\documentclass[sigconf, authorversion=true]{acmart}
\usepackage[english]{babel}
\usepackage[utf8]{inputenc}
\usepackage[T1]{fontenc}

\usepackage{flushend}

\usepackage{hyperref} 
\hypersetup{
    pdfauthor={Adam Krafczyk, Sascha El-Sharkawy, and Klaus Schmid},
    pdftitle={Title},
    pdfsubject={22st International Conference on Software Product Line},
    pdfkeywords={Software product lines; Satisfiability; Integer-based expressions; Propositional logic; Variability management; Reverse engineering},
    pdfpagemode={UseOutlines},
    pdfproducer={LaTeX},
    colorlinks=true,
    linkcolor=black,          
    citecolor=black,          
    filecolor=black,          
    urlcolor=black            
}

\usepackage{listings}
\definecolor{cEclipsePurple}{rgb}{0.5,0,0.35}
\lstdefinelanguage{cWithPre} {
  language=C,
  moredelim = [l][\bfseries\color{cEclipsePurple}]{\#}
}

\usepackage{ifthen}
\newboolean{showToDos}
\setboolean{showToDos}{false} 
\ifshowToDos
  \usepackage{marginnote} 
  \let\marginpar\marginnote
  
  \newcommand{\ToDo}[1]{\noindent\fcolorbox{black}{yellow!20!white}{\parbox{.975\columnwidth}{#1}}}
  \newcommand{\SubjectedToDo}[2]{\ToDo{\textbf{#1:}#2}}
  \newcommand{\ToDoInline}[1]{\fcolorbox{black}{yellow!20!white}{#1}}
  \newcommand{\Outline}[1]{\color{blue}{#1}\color{black}}
  \newcommand{\Problem}[1]{\color{red}{#1}\color{black}}
  \newcommand{\secSize}[1]{[#1]}
  
  \usepackage[textsize=small,colorinlistoftodos]{todonotes}
  \newcommand{\userComment}[2]{\todo[color=#1!20,linecolor=#1!60,bordercolor=#1!60]{#2}}
  \newcommand{\ks}[1]{\userComment{orange}{\textbf{KS:} #1}}
  \newcommand{\se}[1]{\userComment{red}{\textbf{SE:} #1}}
  \newcommand{\ak}[1]{\userComment{blue}{\textbf{AK:} #1}}
  \newcommand{\sd}[1]{\userComment{green}{\textbf{SD:} #1}}

  \usepackage{pifont}
  
\else
  \newcommand{\ToDo}[1]{}
  \newcommand{\SubjectedToDo}[2]{}
  \newcommand{\ToDoInline}[1]{}
  \newcommand{\Outline}[1]{}
  \newcommand{\Problem}[1]{}
  \newcommand{\secSize}[1]{}
  
  \newcommand{\ks}[1]{}
  \newcommand{\se}[1]{}
  \newcommand{\ak}[1]{}
  \newcommand{\sd}[1]{}
\fi

\newcommand{\valFunc}{\sigma}
\newcommand{\emptyVal}{\epsilon}

\begin{document}
\title[Converting Integer-Based Variability to Propositional Logic]{Reverse Engineering Code Dependencies: Converting Integer-Based Variability to Propositional Logic}



\author[A.\ Krafczyk]{Adam Krafczyk}
\affiliation{
  \institution{University of Hildesheim}
  \streetaddress{Universitätsplatz 1}
  \city{31141 Hildesheim} 
  \state{Germany} 
}
\email{adam@sse.uni-hildesheim.de}

\author[S.\ El-Sharkawy]{Sascha El-Sharkawy}
\affiliation{
  \institution{University of Hildesheim}
  \streetaddress{Universitätsplatz 1}
  \city{31141 Hildesheim} 
  \state{Germany} 
}
\email{elscha@sse.uni-hildesheim.de}

\author[K.\ Schmid]{Klaus Schmid}
\affiliation{
  \institution{University of Hildesheim}
  \streetaddress{Universitätsplatz 1}
  \city{31141 Hildesheim} 
  \state{Germany} 
}
\email{schmid@sse.uni-hildesheim.de}


\begin{abstract}

A number of SAT-based analysis concepts and tools for software product lines exist, that extract code dependencies in propositional logic from the source code assets of the product line. On these extracted conditions, SAT-solvers are used to reason about the variability. However, in practice, a lot of software product lines use integer-based variability. The variability variables hold integer values, and integer operators are used in the conditions. Most existing analysis tools can not handle this kind of variability; they expect pure Boolean conditions.

This paper introduces an approach to convert integer-based variability conditions to propositional logic. Running this approach as a preparation on an integer-based product line allows the existing SAT-based analyses to work without any modifications. The pure Boolean formulas, that our approach builds as a replacement for the integer-based conditions, are mostly equivalent to the original conditions with respect to satisfiability. Our approach was motivated by and implemented in the context of a real-world industrial case-study, where such a preparation was necessary to analyze the variability. 

Our contribution is an approach to convert conditions, that use integer variables, into propositional formulas, to enable easy usage of SAT-solvers on the result. It works well on restricted variables (i.e.\ variables with a small range of allowed values); unrestricted integer variables are handled less exact, but still retain useful variability information.





\end{abstract}

%
%

\ccsdesc[500]{Theory of computation~Equational logic and rewriting}
\ccsdesc[500]{Software and its engineering~Software product lines}
\ccsdesc[100]{Software and its engineering~Software reverse engineering}


\keywords{Software product lines; Satisfiability; Integer-based expressions; Propositional logic; Variability management; Reverse engineering}


\setcopyright{acmlicensed}

\acmDOI{10.1145/3236405.3237202}

\acmISBN{978-1-4503-5945-0/18/09}

\acmConference[SPLC'18]{22nd International Conference on Software Product Line}{September 10--14, 2018}{Gothenburg, Sweden}
\acmYear{2018}
\copyrightyear{2018}

\acmPrice{15.00}

\maketitle

\section{Introduction \secSize{1p}}
\label{sec:introduction}

Many software product lines use the C-preprocessor for implementing variability in their source code \cite{PreprocessorVariability, VariabilityAnalysisFortyCppSpls}. The \texttt{\#if}-statement is used to conditionally compile source code parts. The expressions use variability variables to determine whether the following code lines should be included in the product for a given configuration. Typically, such product lines use mostly Boolean variables. A well known example for this kind of code variability is the Linux Kernel. Most of its variability variables are Boolean or tristate. Tristates allow three different values, and are implemented using pairs of Boolean variables in the C-preprocessor.

There exist a lot of  family-based analysis \cite{FamilyBasedAnalysis} approaches and tools for these kinds of product lines. Undertaker finds code blocks that can never be (un-)selected \cite{UndertakerPaper, undertaker_tool}. TypeChef, among other things, finds type errors across all configurations at once \cite{TypechefPaper, typechef_tool}. The feature-effect approach defines Boolean formulas that represent under which condition a given variability variable has an effect on the final product \cite{FeatureEffects}. The configuration mismatch analysis builds on that and finds constraint mismatches between the code and the variability model \cite{ConfigMismatches}. All of these approaches have in common that they use SAT-solvers (or techniques similar to SAT) to reason about the variability in the source code. Because of this, they only work with code dependencies in propositional logic; all variability variables are expected to be Booleans, and the \texttt{\#if}-conditions in the code may only contain Boolean operations.

However, there are also software product lines that are not limited to propositional logic in their variability. For example, an industrial product line we analyzed uses integer-based variability~\cite{BoschPaper}. Each variability variable holds an integer value, out of a defined (typically small) range of allowed values. The \texttt{\#if}-conditions in the code then use the integer arithmetic and comparison operators that the C-preprocessor defines. None of the existing analysis tools mentioned above can handle such variability conditions, because they are built on pure SAT-based approaches. However, we still wanted to use some of the analysis approaches that they offer on the integer-based product line.

To deal with this incompatibility, we developed and implemented an approach to convert the variability in the source code of the product line to a format that is suitable for the existing analysis tools. Our tool creates a copy of the source files in the product line, and replaces all integer-based \texttt{\#if}-conditions with propositional formulas. This allows easy usage of SAT-solvers on the converted code conditions. As a result, all existing analysis tools are able to extract variability information from the modified source code files, without any modifications of the tools. Our tool is implemented in the context of the ITEA3 project $\text{REVaMP}^2$, as an open-source plugin for the KernelHaven analysis framework \cite{KernelHavenPaper, KernelHavenDemoPaper, nonbooleanutils_tool}.

This paper introduces the approach we developed to transform integer-based variability conditions to propositional logic. It reverse engineers the code dependencies from an existing product line, and transforms them into a format that is understood by existing, SAT-based analysis tools. This is based on the industrial use case we analyzed. The variability variables all hold integer values, and most of them have a (typically small) range of allowed values. These ranges are an input to the transformation process and are, for example, defined in a variability model of the product line. The execution time of the transformation process scales linearly with the size of the product line to transform. The overhead is generally insignificant, compared to the following analysis steps.

Some corner cases can not be covered by our approach; thus the resulting propositional formulas are sometimes not completely equisatisfiable, compared to the original integer-based ones. However, we implemented a fallback strategy that produces less accurate results, but still retains useful variability information in these cases. In practice, the resulting formulas are good enough for the analysis tools mentioned above to produce reasonable results.

The remainder of this paper is structured as follows: Section~\ref{sec:related_work} presents work related to integer-based variability. Section \ref{sec:concept} introduces the basic concept of our transformation approach. Based on this, Section \ref{sec:realization} describes the implementation of our approach in more detail and walks through a full example transformation. Section \ref{sec:limitations} discusses the limitations of our approach, while Section \ref{sec:evaluation} evaluates the implementation. Finally, Section \ref{sec:summary} summarizes this paper and shows possible future work.

\section{Related Work \secSize{0.75p}}
\label{sec:related_work}

Existing research shows that integer-based (or more broadly: non-Boolean) variability is used in real-world software product lines. Passos et al.\ studied the non-Boolean variability used in the eCos product line \cite{nonBooleanEcos}. They specifically study the kinds of constraints that non-Boolean variables are used in, and describe the challenges that this kind of variability poses for analyses. Berger et al.\ studied real-world variability modeling techniques \cite{varModelStudy}. All the product lines they studied use integers as variability variables, although the percentage of variability that is integer-based varies. They found that academic analyses often make simplifying assumptions about the structure of variability. In contrast to that, our approach  presented in this paper was driven by a real-world industrial use-case, that uses integer-based variability \cite{BoschPaper}. Thus, it does not include the common simplifications found in some academic concepts.

Analyses that run on product lines that use integer-based variability have to use strategies to deal with the integer variables and constraints. This could be done by employing a solver that can directly work with such constraints. Barrett et al.\ developed a first-order logic solver that incrementally translates parts of the expression to SAT, while solving it \cite{FirstToPropSolving}. Satisfiability modulo theory (SMT) solvers, for example the Z3 solver by Microsoft \cite{Z3_Paper}, handle first-order logic conditions directly. Such a solver does not require a translation to SAT. Xiong et al.\ use Z3 for their approach to fix ill-formed configurations that contain integer variables \cite{rangeCalculation}. An ill-formed configuration violates one or more of the constraints imposed by the variability model. Their approach calculates valid ranges for integer variables, that adhere to these violated constraints. It works directly on the integer-based constraint, without a translation to SAT.

The existing SAT-based analysis approaches and tools, introduced in Section \ref{sec:introduction}, could use such a solver to handle integer-based variability. However, this would require to modify the tools; they need to be extended to parse and represent the integer-based conditions, and their SAT-solver needs to be swapped for one of the solvers mentioned above. In contrast, our approach works as a preparation for the product line, and no modification is required to the SAT-based analysis tools. Our approach takes advantage of the fact that in the analyzed product line, most of the integer variables have only a limited range of allowed values. Additionally, we expect the runtime of a more powerful solver to be much higher than the runtime of our preparation approach plus the simple SAT-solvers. 



\section{Concept \secSize{1.25p}}
\label{sec:concept}

This section introduces our concept for converting integer-based conditions to propositional formulas. First, we describe the structure of the integer-based conditions of the product line. Then we explain why existing analysis tools can not work with these conditions directly. Finally, we introduce our approach that solves the incompatibility between the integer-based conditions and the existing, propositional logic-based analysis tools.

The source code in product lines that we want to analyze contains integer-based variability conditions in the C-preprocessor. The variables used in the conditions hold integer values, and have a (typically small) range of allowed values. All integer arithmetic operations ($+$, $-$, $*$, $/$, $\%$, $\&$, $|$, $\verb!^!$, $\sim$) and integer comparison operators ($==$, $!=$, $>$, $>=$, $<$, $<=$) that the C-preprocessor defines are used in the conditions. Additionally, the Boolean operators ($\&\&$, $||$, $!$) and the \texttt{defined} function are used. The \texttt{defined(VAR)} function returns whether any value has been set for the variable. For instance, a condition from such a product line may look like this:
\begin{equation}
	\label{eq:integerExample}
	\texttt{\#if\ (VAR\_A * 2 > VAR\_B)\ ||\ defined(VAR\_C)}
\end{equation}
with the variability model specifying the possible ranges for the integer variables: $\texttt{VAR\_A} \in \{1, 2, 3\}$ and $\texttt{VAR\_B} \in \{5, 6\}$.

The existing tools for analyzing C-preprocessor based variability, introduced in Section \ref{sec:introduction}, can only handle conditions in propositional logic. This is because they use SAT-based approaches to reason about variability. Only Boolean operators ($\&\&$, $||$, $!$) and Boolean variables are allowed in the variability conditions that they extract. The convention for Boolean variables in the C-preprocessor is to use the \texttt{defined} function on a variable, and denoting the two possible states by either defining or not defining the variable. An example of such a product line is the Linux Kernel, which is also a common target for research analysis tools \cite{UndertakerPaper}. The existing analysis tools expect the analyzed code to follow this convention for pure Boolean conditions; they can not handle integer-based conditions, where the value that a variable holds is important. Additionally, they often simply can not parse the integer operations that are used in the conditions.

The goal of our approach is to construct conditions in propositional logic, that are, with respect to satisfiability, mostly equal to the integer-based conditions in the source code. Thus, after preparing the code artifacts with our approach, all the existing tools can be used on the previously integer-based product line. The propositional replacements hold mostly the same variability information as the integer-based conditions, with respect to the satisfiability of the code dependencies. This ensures that the results produced by the SAT-based analysis approaches are reasonable and useful for the analyzed product line.

Our approach introduces Boolean variables for each possible value of the integer variables. This is feasible, because most of the integer variables have only a small range of allowed values. For each integer-based condition, we calculate which combinations of values satisfy this condition. From this, a propositional formula using the introduced Boolean variables is created, that reflects these combinations. Additionally, we introduce another Boolean variable for each integer variable, that denotes whether the variable is set to any value, or whether it is undefined. This is used for resolving \texttt{defined} calls, and as a fallback for cases that our approach can not calculate exact solutions for (see Section \ref{sec:limitations}).

\ToDo{Diagram: variable ranges + constants + input code $\rightarrow$ output code}

More formally, let $V$ be the set of all integer-based variability variables, $R: V \rightarrow \mathcal{P}(\mathbb{Z})$ a function that defines the range of allowed values for each variable, and $B$ a set of Boolean variables. We then introduce a function $$\valFunc:\ V\ \times\ (\mathbb{Z}\ \cup\ \{\emptyVal\})\ \rightarrow\ B$$ which injectively maps a variable and one possible value of it to a Boolean variable. $\emptyVal$ in place of a value maps to the Boolean variable that denotes whether the integer variable is defined or not (i.e. whether it is set to \textit{any} value). For instance, consider the variable \texttt{VAR\_A} with $R(\texttt{VAR\_A}) = \{1, 2, 3\}$. \texttt{VAR\_A} may be set to either 1, 2, or 3. $\valFunc(\texttt{VAR\_A}, \emptyVal)$ returns the Boolean variable that denotes whether \texttt{VAR\_A} is defined. $\valFunc(\texttt{VAR\_A}, 1)$ returns the Boolean variable that denotes that \texttt{VAR\_A} is set to 1. In practice, such variable names may look like this: $\valFunc(\texttt{VAR\_A}, 1) = \texttt{VAR\_A\_eq\_1}$, $\valFunc(\texttt{VAR\_A}, \emptyVal) = \texttt{VAR\_A}$. The exact naming scheme depends on the context that the approach is applied in; it has to ensure that no name collisions occur, and that the names are valid identifiers for the C-preprocessor.

There are two constraints for the introduced Boolean variables, which are not explicitly modeled. These have to be manually considered when interpreting the result of any analysis done on the propositional formulas:
\begin{enumerate}
	\item The Boolean variables for the possible values of an integer variable are mutually exclusive: $$\forall v \in V,\ \forall i, j \in R(v), i \neq j |\ \valFunc(v, i) \implies \neg \valFunc(v, j)$$
	\item The $\emptyVal$ Boolean variable is true if and only if a value is set for the integer variable, that is if any of the Boolean variables denoting the possible values is true: $$\forall v \in V\ |\ \valFunc(v, \emptyVal) \iff \bigvee_{i \in R(v)} \valFunc(v, i)$$
\end{enumerate}

These newly introduced Boolean variables are used to replace the integer-based (sub-)expressions in the conditions. We calculate which combinations of allowed values fulfill the integer-based (sub-)expression, and build a Boolean formula that is equally satisfiable to this. For the example condition shown in Formula \ref{eq:integerExample}, the propositional replacement would look like this: $$\texttt{\#if\ (VAR\_A\_eq\_3\ \&\&\ VAR\_B\_eq\_5)\ ||\ VAR\_C}$$ (the \texttt{defined} calls around each of these variables has been left out for brevity). The integer sub-expression \texttt{VAR\_A * 2 > VAR\_B} is only fulfilled by the combination \texttt{VAR\_A=3} (with $\valFunc(\texttt{VAR\_A}, 3) = VAR\_A\_eq\_3$) and \texttt{VAR\_B=5} (with $\valFunc(\texttt{VAR\_B}, 5) = VAR\_B\_eq\_5$)\ToDoInline{formatting}. Thus, the replacement for this sub-expression is a Boolean formula for this combination of values. The sub-expression \texttt{defined(VAR\_C)} is replaced with $\valFunc(\texttt{VAR\_C}, \emptyVal) = VAR\_C$.

In many cases, the resulting propositional formula is much larger, compared to the original integer-based one. The goal, however, is not to produce small or readable conditions, but to provide input data for analysis tools. Thus, it is not a primary concern to keep the resulting propositional formulas concise.

A special case are integer variables that have no restriction on the allowed values. The ``allowed values'' for such a variable are for example the whole range of 32 bit integer variables. It is not possible (or at least not feasible) to introduce Boolean variables for each of the possible values. In this case, we take a less exact approach to still be able to handle these variables: we only introduce the single Boolean variable $\valFunc(var, \emptyVal)$. Then, in each condition where the unrestricted variable appears, we use this Boolean variable, no matter which actual value of the variable would fulfill the condition. This way, we still have the variability information that the given condition \textit{somehow} depends on the unrestricted variable; however, we lose the information which specific values it depends on. For example, the condition $\texttt{VAR\_A} + 2 > 5$, with \texttt{VAR\_A} as an unrestricted integer variable, is converted to $\valFunc(\texttt{VAR\_A}, \emptyVal)$.

Another special case are variables with only one possible allowed value. These variables are not real variability, since they can not be set to any other value. Thus, we treat them as constants and replace each occurrence of them with their literal value. This makes it easier to calculate the combinations of the other integer variables, that satisfy a constraint. It also removes unnecessary variables from the propositional formulas, and thus reduces the unnecessary complexity of the output.

\section{Implementation \secSize{3p}}
\label{sec:realization}

This section describes the implementation of our approach. First, the general steps for converting a condition are explained. Then, the transformation of the integer (sub-)expressions is described in detail in Section \ref{sec:transformation}. This is followed by Section \ref{sec:full_example} with an example of a complete conversion of a code condition. 

Our tool reads through all source files and replaces the C-pre\-processor conditions in them with a propositional formula. The result is a copy of the source files with all the C-preprocessor conditions transformed into a purely Boolean form. This allows existing analysis tools, which are based on propositional logic, to work with these files. Each of the C-preprocessor conditions found in the source files is converted in the following steps:


\begin{enumerate}
	\item Parse the condition into an abstract syntax tree (AST)
	\item Replace constants with their literal value
	\item Walk bottom-up through the AST to replace integer-parts
	\item Convert AST back to a C-preprocessor condition string
\end{enumerate}

The first step is straightforward parsing of the C-preprocessor condition. All the subsequent steps will work on the AST that is produced by this. In the second step, each constant that has only one allowed value is replaced by its literal value. This removes constant values, which do not need to be considered when analyzing variability.

The third step is the main part of the conversion to a propositional form. The goal is to convert all integer-parts of the conditions to pure Boolean ones, that have equivalent satisfiability. The strategy used here is to walk bottom-up through the AST and apply a number of rules on the AST that define how the integer operations are converted. The following section will explain these rules in detail.

Finally, after the integer parts are eliminated, the AST is converted back into a C-preprocessor condition string. This string is then used as the replacement for the original condition. It will only contain Boolean variables and operators, so that existing tools for propositional logic can handle it.

\ToDo{
\begin{itemize}
	\item Diagram: show example AST, and highlight integer and boolean parts
\end{itemize}
}

\subsection{Transformation of Integer Expressions}
\label{sec:transformation}

This section describes how the integer-based (sub-)expressions are converted to Boolean formulas. In the AST, the highest operator of an integer (sub-)expression is a comparison operator\footnote{Sometimes there are no explicit comparison operators to convert integer expressions to Boolean values. In this case, a \texttt{!= 0} comparison can be assumed, since all integer values except 0 are defined to be \texttt{true} in the C-preprocessor.}. On the left and right side of this comparison, there are literals, variables, or arithmetic operations combining both. The general idea is to find all possible combinations of allowed values for the variables on the left and right side that fulfill the comparison operator. These values are then transformed into Boolean variables using the $\valFunc$ function. A propositional formula is constructed from them, that is satisfiable for all combinations that fulfill the comparison.

Integer arithmetic operations are evaluated bottom-up, so that eventually the comparison operation can be evaluated on the resulting values. When resolving the arithmetic operations, they are applied on all possible values of a variable. This results in a set of values, instead of a single result value for the operation. This is needed, because we want to find all possible values that fulfill the comparison at once.

It is also not enough to just compute the results of integer arithmetic operations on variables. When a result value that fulfills the comparison operator is found, the original value of the variable that led to this result value is required to construct the Boolean variable using the $\valFunc$ function. We call the result of arithmetic operations the \textit{current value} and the initial value of the allowed range of the variable that this result stems from the \textit{original value}. For instance, consider the integer expression $\texttt{VAR\_A} + 1 == 2$. When evaluating the $+$ operation, the \textit{original} value 1 of \texttt{VAR\_A} led to the \textit{current} value 2. When resolving the equality operator, the \textit{current} value 2 for the left side fulfills this comparison. Thus, the Boolean variable of the \textit{original} value that led to this is constructed: $\valFunc(\texttt{VAR\_A}, 1)$.

The following sub-sections describe the different evaluation rules in detail. The rules are applied on the AST, based on which integer operator is used on which input types.

\begin{itemize}
	\item \textit{comparison operator} refers to integer comparison operators ($==$, $!=$, $<$, $<=$, $>$, $>=$)
	\item \textit{arithmetic operator} refers to integer arithmetic operators ($+$, $-$, $*$, $/$, $\%$, $\&$, $|$, $\verb!^!$, $\sim$)
	\item \textit{literal} refers to literal integer values
	\item \textit{variable} refers to integer variables, with a defined range of allowed values
\end{itemize}

\subsubsection{Arithmetic Operator on two Literals}

For integer arithmetic operations on two literal values, the result is simply calculated.

\begin{figure}[H]
\centering
\includegraphics[scale=0.3]{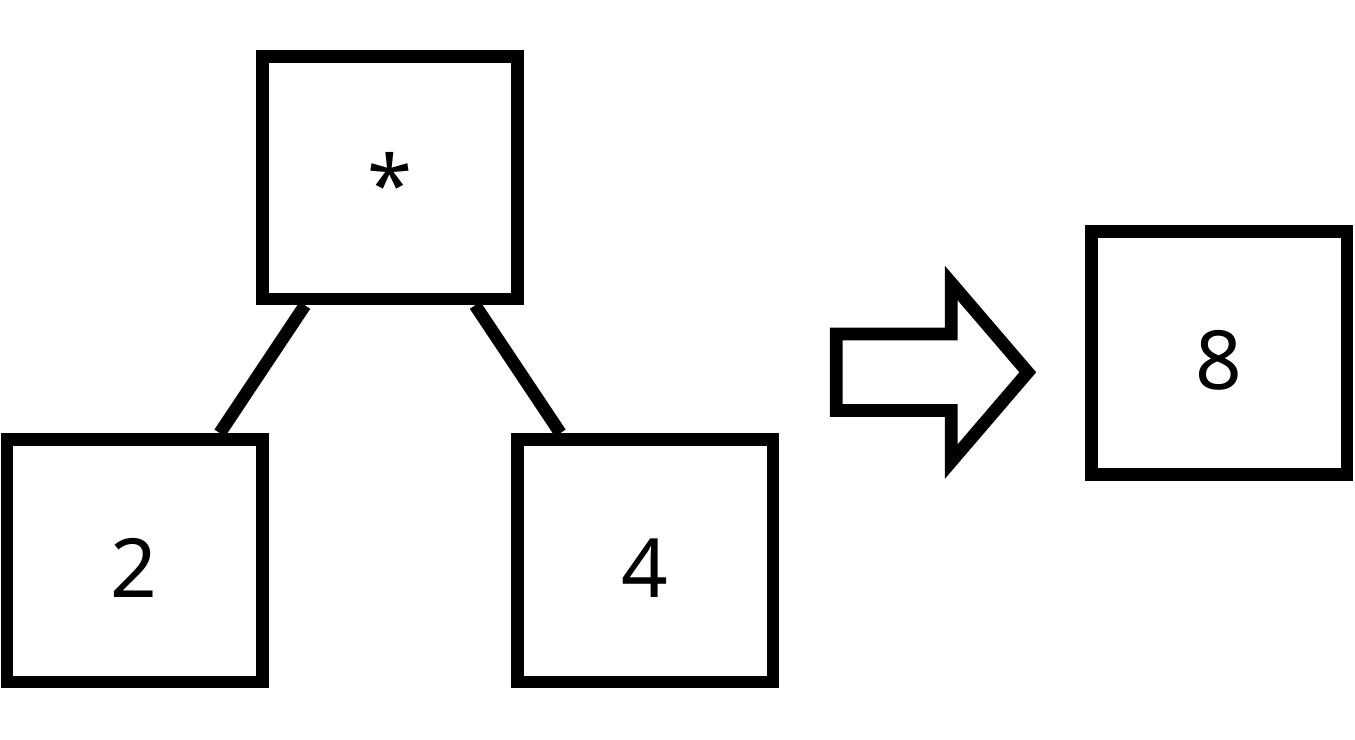}
\end{figure}

In this figure, the expression \texttt{2 * 4} results in the literal value \texttt{8}.

\subsubsection{Comparison Operator on two Literals}

For comparison operations on two literal values, the resulting Boolean constant is simply calculated. Neither side of the comparison contains any integer variables, thus a single Boolean literal can express the satisfiability of this (sub-)expression.

\subsubsection{Arithmetic Operator on Literal and Variable}

For integer arithmetic operations on variables, the operation is calculated on each of the allowed values. A set of tuples is stored in the variable, which contains for each \textit{original} value, the \textit{currently} computed value. All arithmetic operations on this variable will always update the \textit{current} value. When resolving the variable to a Boolean formula later on, it is important which \textit{original} value led to the currently computed one.

\begin{figure}[H]
\centering
\includegraphics[scale=0.3]{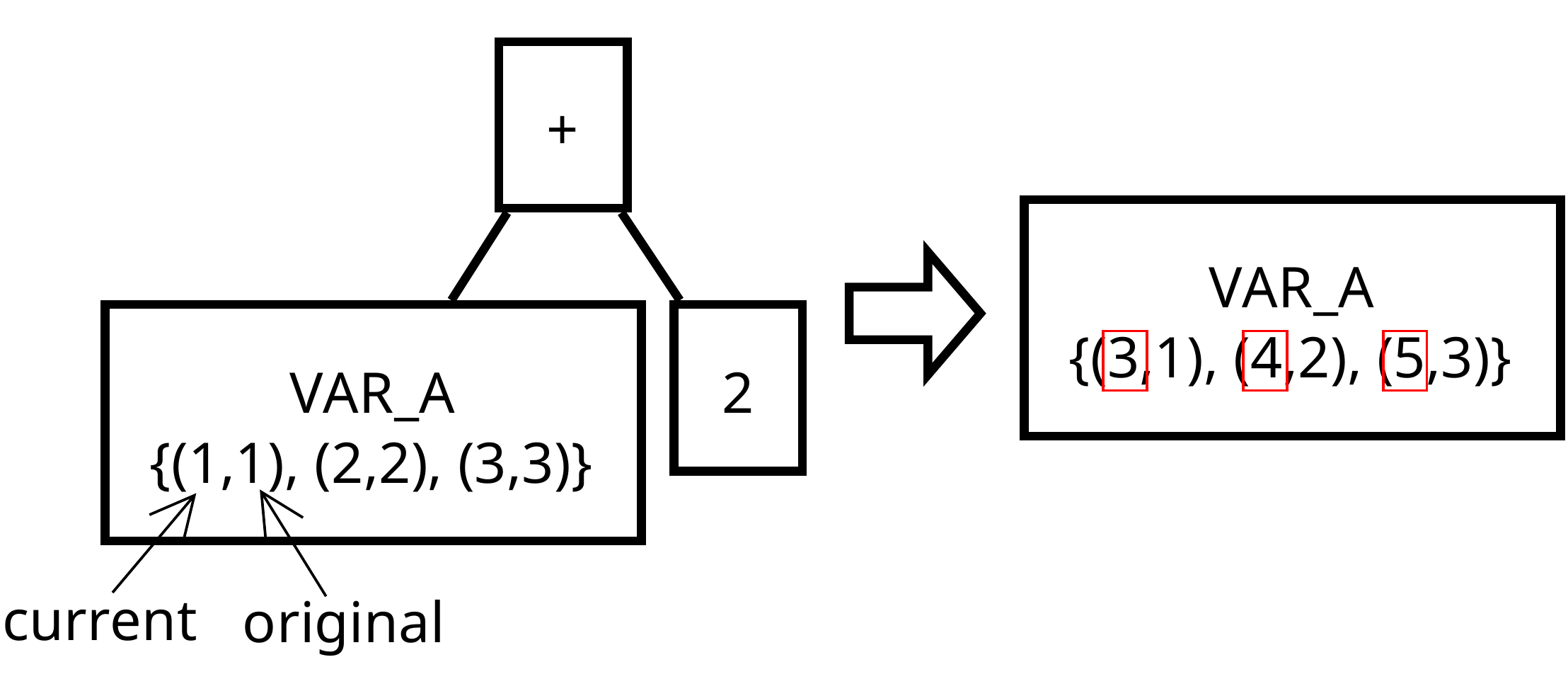}
\end{figure}

In this figure, the literal value 2 is added to \texttt{VAR\_A}. \texttt{VAR\_A} has three possible values: 1, 2, and 3. For each of these possible values, the operation is computed and the result stored in the first component in the tuple. The second component is not modified; it contains the \textit{original} value of \texttt{VAR\_A} that led to the \textit{currently} computed one. For example, in the second tuple of the result, the \textit{original} value 2 of \texttt{VAR\_A} led to the \textit{current} value of 4 (via the addition of 2).

\subsubsection{Comparison Operator on Variable and Literal}

A comparison of a variable and an integer literal is resolved to a propositional formula, which contains all possible \textit{original} values that satisfy the comparison. The comparison is computed on all the \textit{current} values stored in the variable. For each \textit{current} value that satisfies the comparison, the corresponding \textit{original} value of the variable is turned into a Boolean variable via the $\valFunc$ function. All these Boolean variables that fulfill the comparison are then combined with a Boolean disjunction operator.

\begin{figure}[H]
\centering
\includegraphics[scale=0.3]{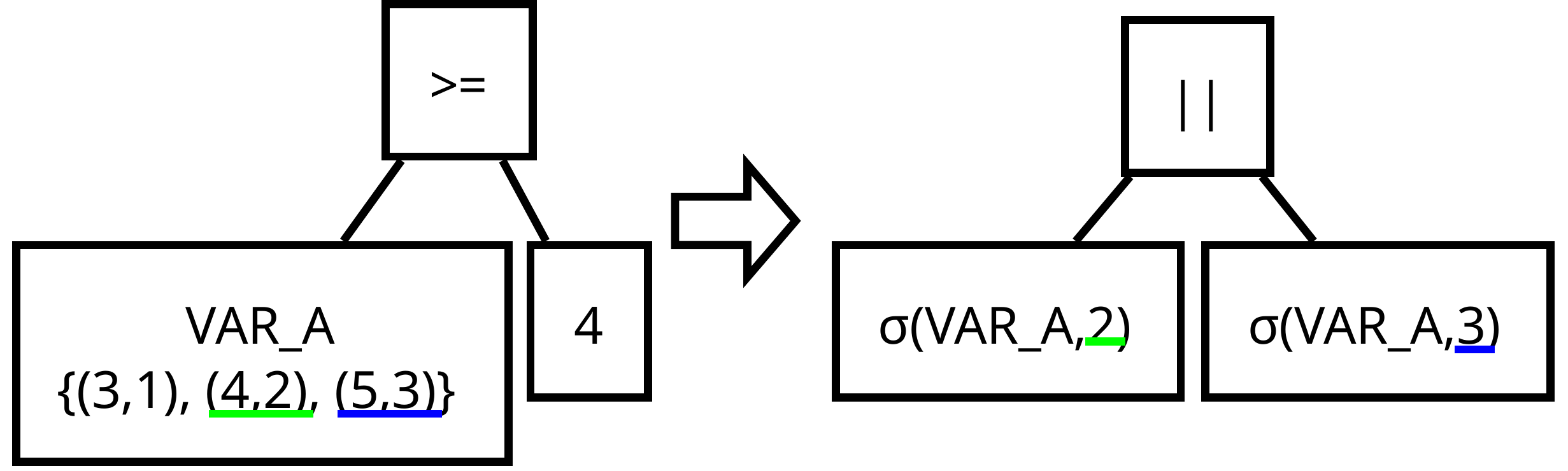}
\end{figure}

In this figure, \texttt{VAR\_A} is compared with the literal 4 with a ``greater than'' comparison operator. In this example, the \textit{current} and \textit{original} values in the tuples are different; this is because some previous arithmetic operation on \texttt{VAR\_A} has modified them. This is not always the case (a variable can also be compared without doing arithmetic on it first), but we chose this for illustration purposes, to make it clear that the \textit{current} and \textit{original} values have to be treated differently. Two of the \textit{current} values (the first component of the tuples) of \texttt{VAR\_A} fulfill the comparison: the second and the third tuple. From both these tuples, the \textit{original} values (the second component) are transformed into Boolean variables and combined with a logical disjunction.

\subsubsection{Comparison Operator on two Variables}

A comparison of two variables is resolved to a propositional formula that contains all possible combinations of \textit{original} values that satisfy the comparison. For each pair of the \textit{current} values of the two variables it is checked if they fulfill the comparison operator. For each pair that does fulfill it, the two \textit{original} values of the variables are turned into Boolean variables (via the $\valFunc$ function) and combined with a logical conjunction operator. All of these conjunction terms are then combined with a logical disjunction operator.

\begin{figure}[H]
\centering
\includegraphics[scale=0.3]{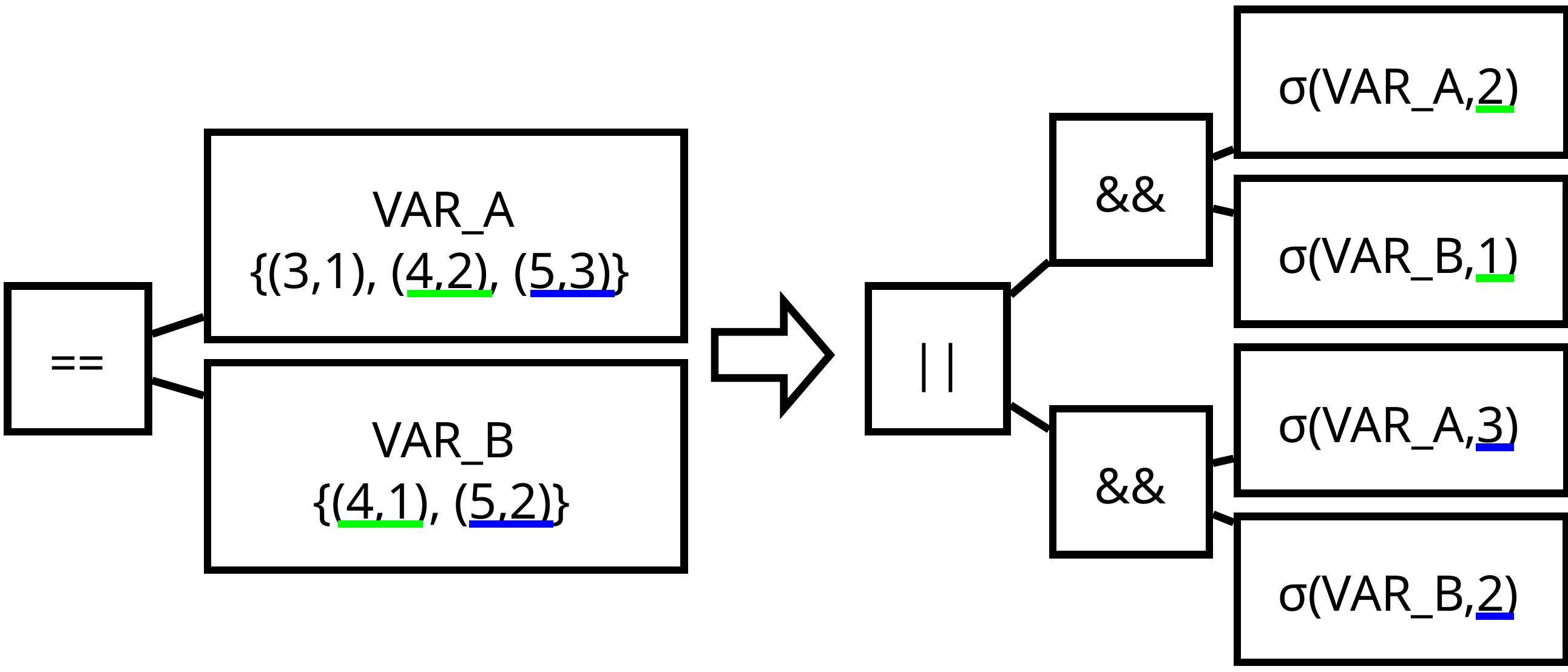}
\end{figure}

In this figure, there are two pairs of tuples that have the same \textit{current} value and thus fulfill the equality operator: the second one from \texttt{VAR\_A} and the first one of \texttt{VAR\_B} both have the value 4, the third one from \texttt{VAR\_A} and the second one from \texttt{VAR\_B} both have the value 5. For the first pair, the \textit{original} value of \texttt{VAR\_A} that led to the \textit{current} value is 2 (the second component in the tuple), while the original value of \texttt{VAR\_B} is 1. Thus, the Boolean representation of this combination is $\valFunc(\texttt{VAR\_A}, 2) \land \valFunc(\texttt{VAR\_B}, 1)$. Similarly, the Boolean representation of the second matching pair is $\valFunc(\texttt{VAR\_A}, 3) \land \valFunc(\texttt{VAR\_B}, 2)$. Since both of these pairs fulfill the equality operator, they are combined with a disjunction operator.

\subsubsection{Arithmetic Operator on two Variables}

For integer arithmetic operations on two variables, the operation is done on each combination of the \textit{current} values of both variables. For each of these calculated values, both of the \textit{original} values of the variables that led to this \textit{current} value are stored. When turning this tuple into a Boolean formula, not a single variable is created (e.g. $\valFunc(\texttt{VAR\_A}, 2)$), but a logical disjunction of the two variables with the \textit{original} values stored in the tuple (e.g. $\valFunc(\texttt{VAR\_A}, 2) \land \valFunc(\texttt{VAR\_B}, 1)$) \ToDoInline{formulation}.

\begin{figure}[H]
\centering
\includegraphics[scale=0.3]{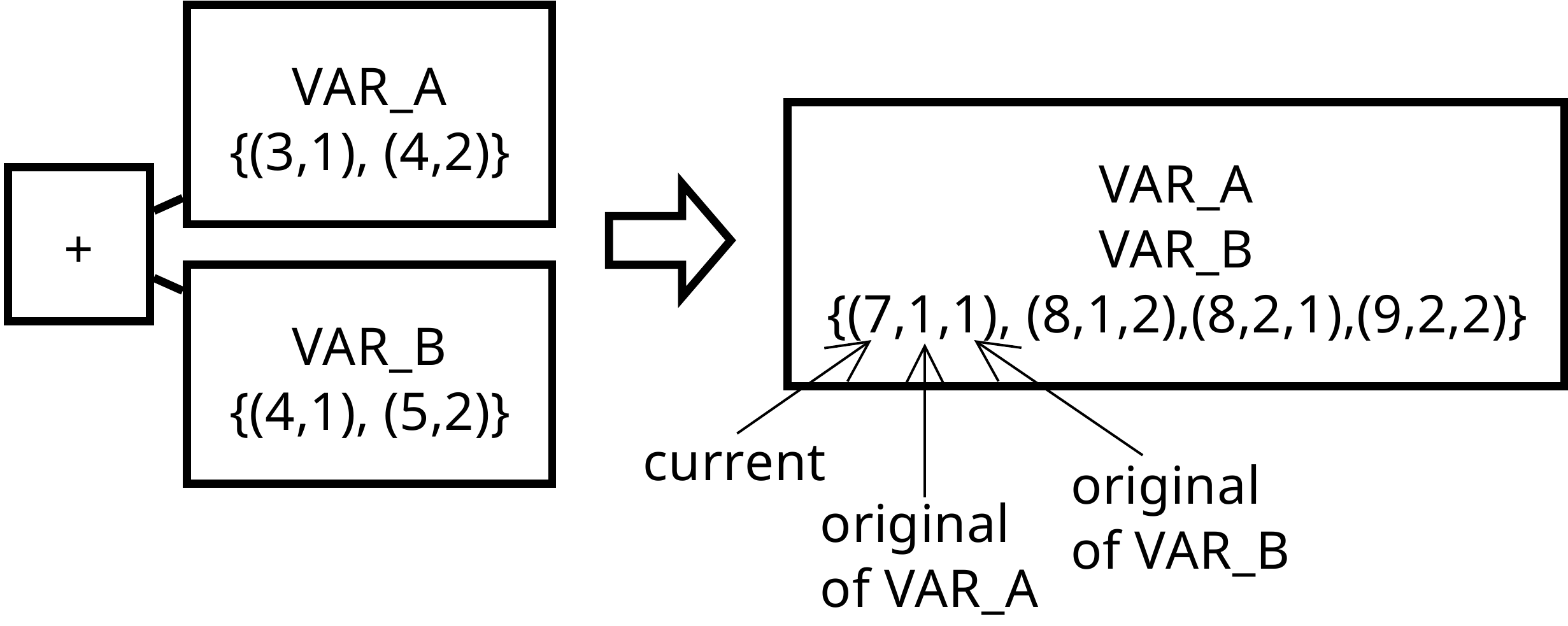}
\end{figure}

In this figure, the two variables \texttt{VAR\_A} and \texttt{VAR\_B}, both with two possible values, are added together. Combining the first tuple of both, results in the first tuple of the result: the \textit{current} values (the first components of the tuples: 3 and 4) are added together, resulting in the new \textit{current} value 7. Then, both of the \textit{original} values (the second components in the tuples) are stored in the result, to indicate which \textit{original} values of \texttt{VAR\_A} and \texttt{VAR\_B} led to the current value of 7. When turning this tuple into a Boolean formula (if the \textit{current} value of this tuple fulfills a comparison later on), then the \textit{original} values of both variables have to be considered: the resulting formula is $\valFunc(\texttt{VAR\_A}, 1) \land \valFunc(\texttt{VAR\_B}, 1)$.

\subsection{Full Example}
\label{sec:full_example}

This section shows an example of a full transformation from an integer-based C-preprocessor condition to a propositional formula. The original condition to convert is: $$\texttt{\#if\ (VAR\_A * CONST\_A > VAR\_B)\ ||\ defined(VAR\_C)}$$ with $R(\texttt{VAR\_A}) = \{1, 2, 3\}$, $R(\texttt{VAR\_B}) = \{5, 6\}$, $R(\texttt{VAR\_C}) = \{0, 1\}$, and $R(\texttt{CONST\_A}) = \{2\}$.

The first step is to parse the condition into an abstract syntax tree (AST):

\begin{figure}[H]
\centering
\includegraphics[scale=0.3]{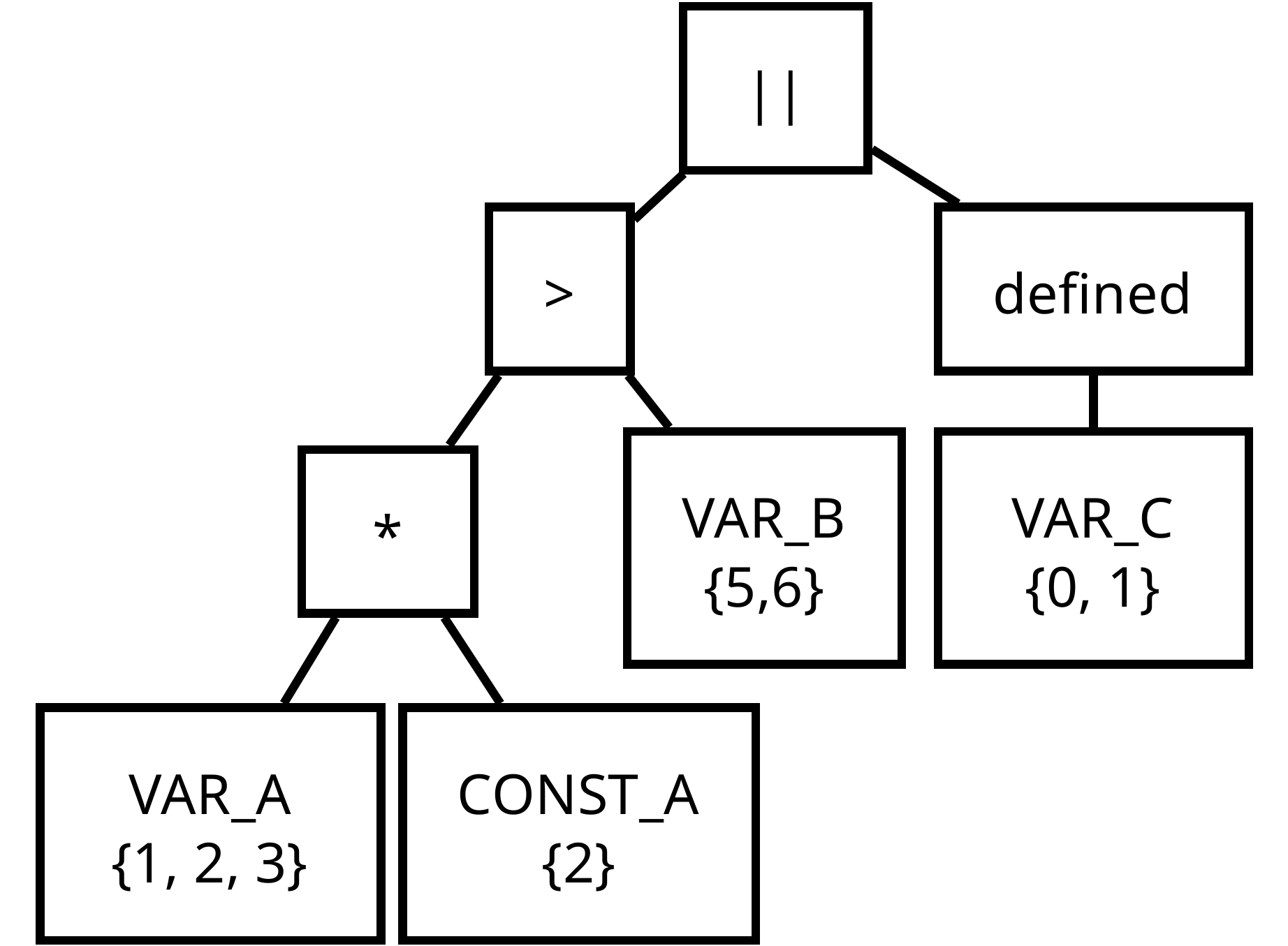}
\end{figure}

The second step replaces the constant \texttt{CONST\_A} with its literal value 2:

\begin{figure}[H]
\centering
\includegraphics[scale=0.3]{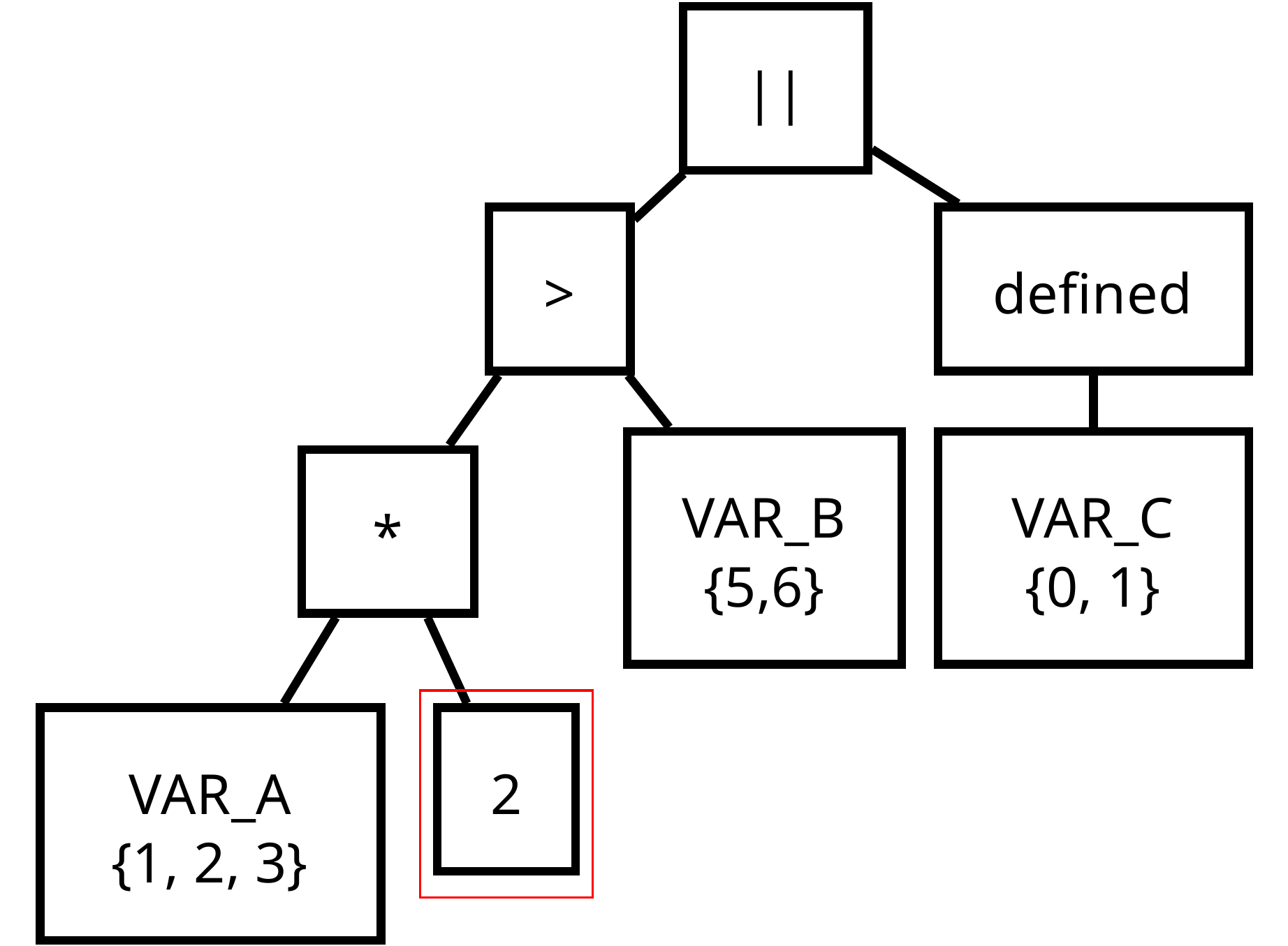}
\end{figure}

The third step is the main part of the conversion. The integer sub-expression on the left side, $\texttt{VAR\_A * 2 > VAR\_B}$, is converted into a propositional formula. First, the multiplication operation is resolved, by multiplying each of the possible values of \texttt{VAR\_A} with the literal value 2:

\begin{figure}[H]
\centering
\includegraphics[scale=0.3]{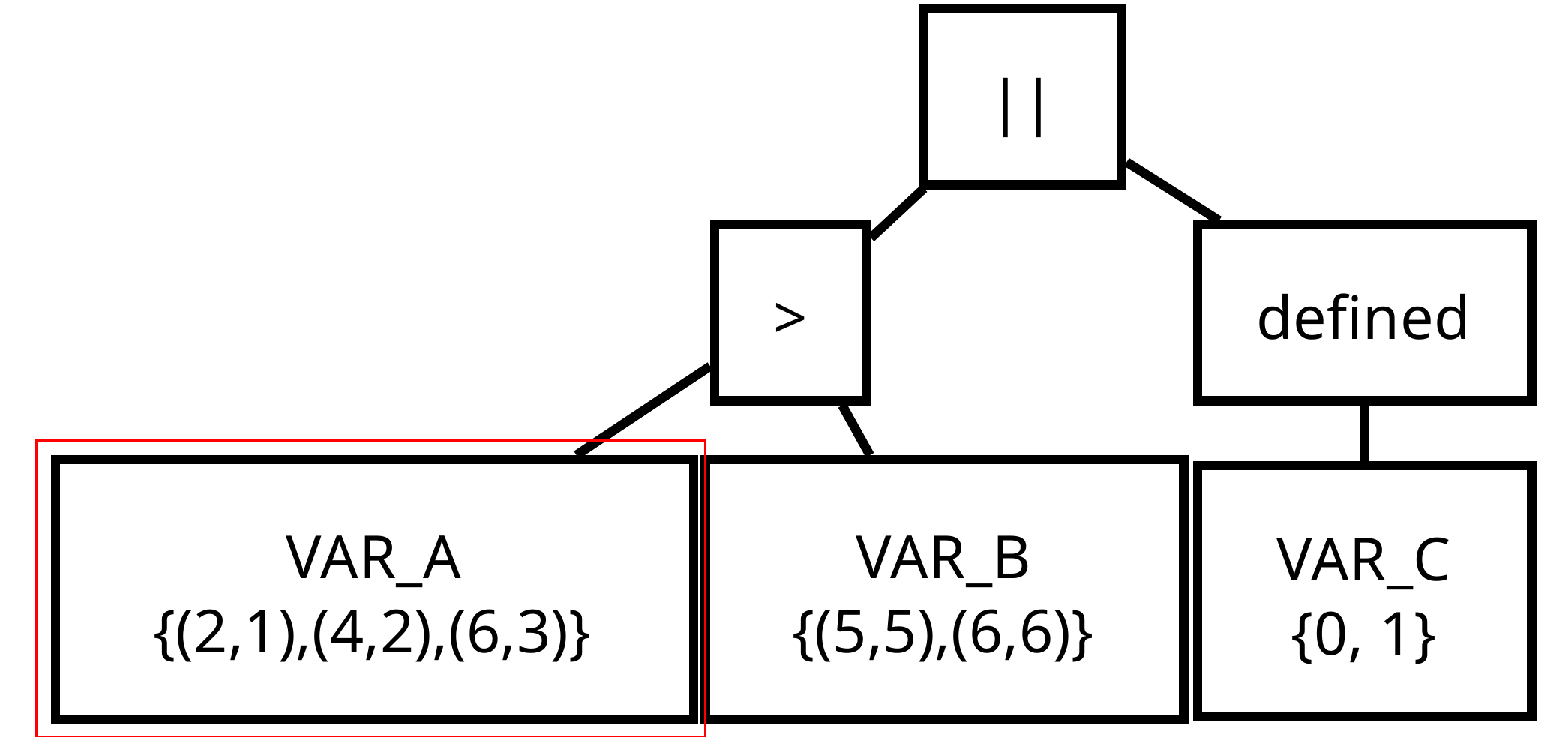}
\end{figure}

Then the comparison operator can be resolved. Only one pair of tuples from \texttt{VAR\_A} and \texttt{VAR\_B} fulfills this: the third tuple of \texttt{VAR\_A} and the first tuple of \texttt{VAR\_B} ($6 > 5$). This is then converted into a propositional formula, that specifies that the original values 3 from \texttt{VAR\_A} and 5 from \texttt{VAR\_B} fulfill this comparison:

\begin{figure}[H]
\centering
\includegraphics[scale=0.3]{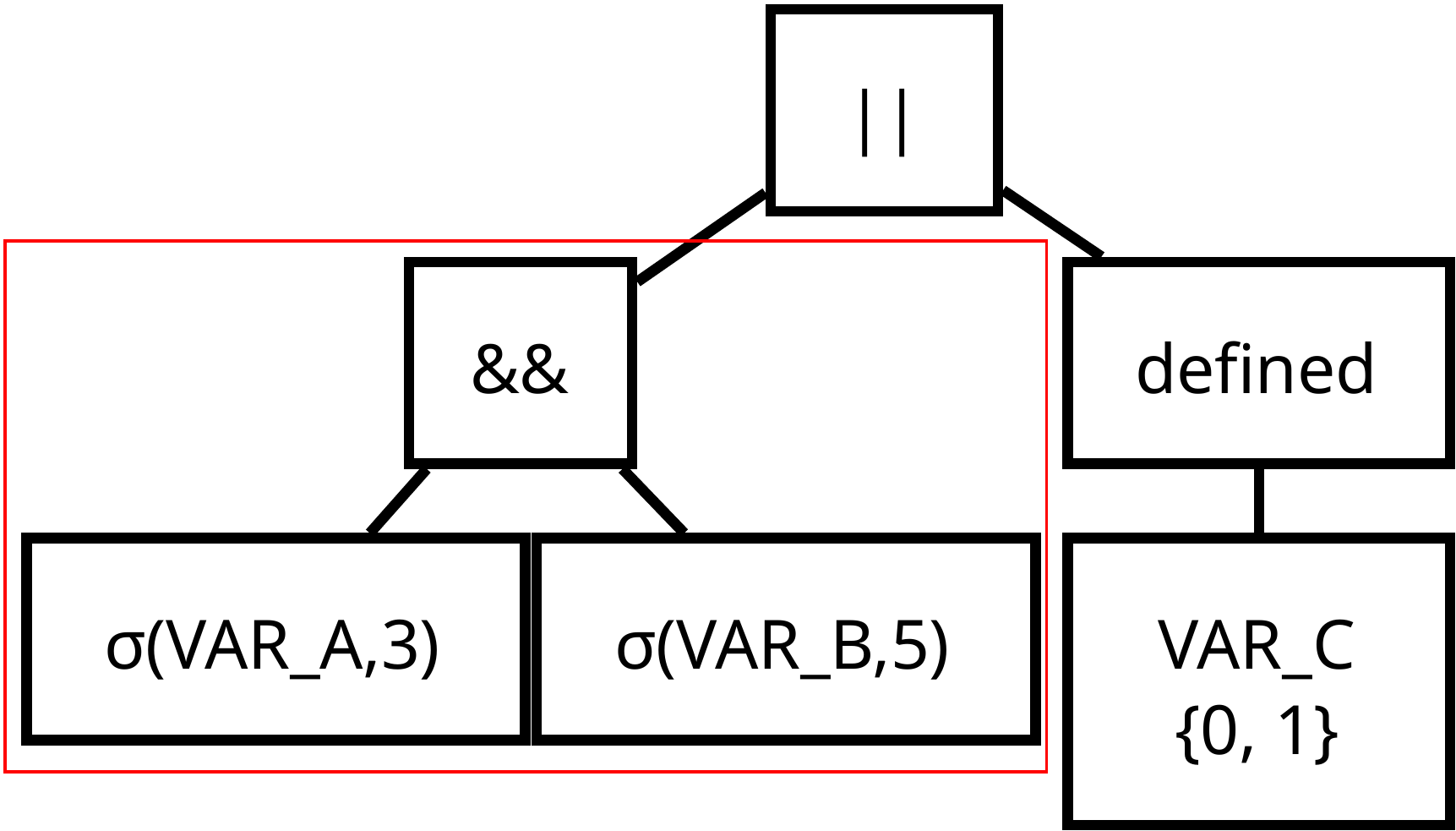}
\end{figure}

The right side of the disjunction operator, \texttt{defined(VAR\_C)}, is replaced with the $\emptyVal$ variable for \texttt{VAR\_C}:

\begin{figure}[H]
\centering
\includegraphics[scale=0.3]{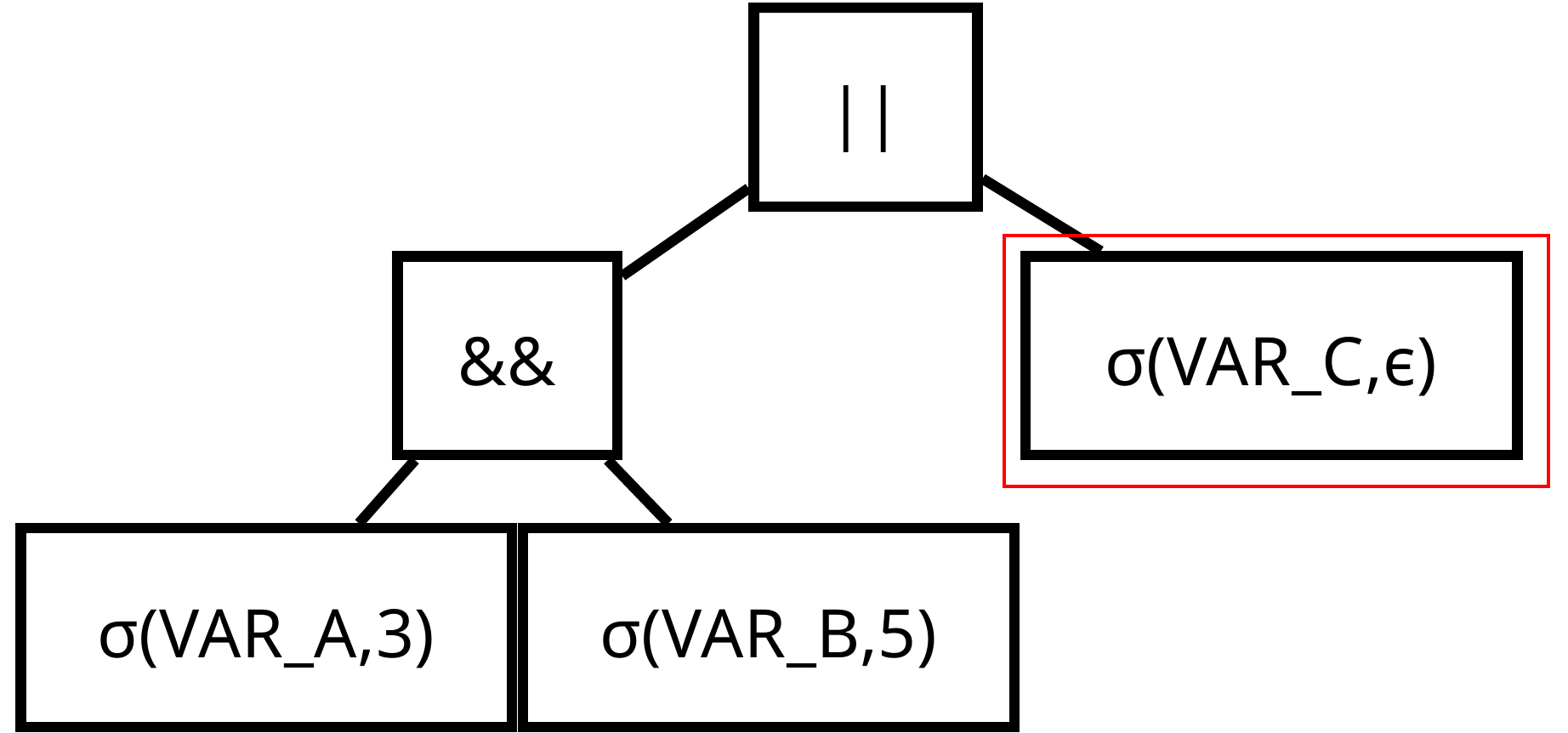}
\end{figure}

The AST now only contains Boolean variables and operators. The fourth and final step is to convert it back into a C-preprocessor condition string (the \texttt{defined} calls around each of the variables have been left out for brevity):
$$\texttt{\#if (VAR\_A\_eq\_3\ \&\&\ VAR\_B\_eq\_5)\ ||\ VAR\_C}$$

\ToDo{Section about unrestricted integer variables?}



\section{Limitations \secSize{0.75p}}
\label{sec:limitations}

This section will discuss the limitations of the approach presented in this paper. Both, the conceptual problems of our approach, and the technical issues of our implementation are discussed.

An important conceptual problem is that unrestricted integer variables are handled not exact. For integer variables that have an infinite (or very large) range of allowed values, our approach does not differentiate between the different values that this variable holds. Only the fact that a condition depends on the variable having a defined value is represented in the replacement propositional formula. For example, the condition $\texttt{VAR\_A} * 2 > 5$, with \texttt{VAR\_A} as an unrestricted integer variable, leads to the replacement condition $\texttt{defined(VAR\_A)}$. This will also lead to some conditions, such as $\texttt{VAR\_A} > 0\ \&\&\ \texttt{VAR\_A} < 0$, to appear satisfiable in the propositional formula.

The impact of this inexact strategy depends on the concrete use-case where our approach is applied. The more unrestricted integer variables are present in the variability model, the more of a problem this becomes. Also the usage of these variables in the conditions needs to be examined: if the unrestricted variables are mixed together with the other integer variables in the conditions, then the inexact results may influence the other variables, too. In contrast, if they are mostly used in separate conditions, then the results for the restricted variables are unaffected and remain exact.

In our industrial use-case, we found that this inexact strategy does not have a large effect on calculating feature-effects \cite{BoschPaper}. The feature-effect analysis builds formulas from the variability conditions in the code, which specify when a certain feature has an effect on the final product. If a condition contains an unrestricted integer variable, the propositional formula for it created by our approach retains the information that the condition \textit{somehow} depends on this unrestricted variable. This is because our approach adds a $\valFunc(\texttt{VAR}, \emptyVal)$ variable in place of the unrestricted variable. This dependency then also appears in the calculation of the feature effects. It is not as exact as the per-value analysis of the restricted variables, however the general dependency is still included.

Another conceptual limitation of our approach is that resulting propositional formulas may be much larger than the original integer-based ones. This is because our approach considers all possible values for integer variables when resolving integer operations on them. When both sides of an operation are integer variables, a pair-wise combination of their possible values is necessary. This leads to a quadratic growth of combinations when multiple variables are combined in a series of arithmetic operations.

Generally, creating long replacement conditions is not a problem, since they are only used as input for further analysis tools. The formulas are not meant to be human-readable. However, in practice, we encountered runtime and memory problems with too large conditions in a very small number of cases.

To circumvent the runtime and memory problems of too large conditions, our tool defines a fixed upper limit for the number of value combinations to consider. When a series of arithmetic operations on variables exceeds this limit, we drop the per-value analysis for this sub-expression. Instead, we fall back to a similar approach used for the infinite integer variables: only $\valFunc(\texttt{VAR}, \emptyVal)$ variables are created for all involved variables. This retains the information, that the condition \textit{somehow} depends on these variables, but the information which concrete values it depends on is lost.

Finally, there are a few minor technical issues with our implementation. These stem from the specifics of the C-preprocessor, and are thus not inherent to our approach itself.
\begin{itemize}
	\item The C-preprocessor has no well-defined data types. This leads to a problem when evaluating the bit-wise negation operator $\sim$, where the concrete type of an integer (bit size, and whether it is signed or unsigned) is important for correct results. However, in practice, this operator is not used much.
	\item Based on the underlying industrial use-case, our tool is only designed to handle integer and Boolean variables in the C-preprocessor conditions. It can not handle string variables, or the string concatenation operator ($\#\#$). When encountering this, our tool will print a warning and skip replacing the condition.
	\item The C-preprocessor allows defining functions (with the \texttt{\#define} statement) that can be used in \texttt{\#if}-statements. Our tool can not interpret these functions; when such a function appears in a condition, our tool will print a warning and skip replacing the condition.
\end{itemize}

\section{Evaluation \secSize{0p}}
\label{sec:evaluation}

We have evaluated the implementation of the approach presented in this paper both in practice and with generated test cases. The generated test cases were used to evaluate the performance of the implementation. Each test case consists of 100 generated C source files with 10 \texttt{\#if}-conditions each. The \texttt{\#if}-conditions are generated in a way that each of the transformation rules described in Section~\ref{sec:transformation} are covered. 5 integer variables are used throughout these conditions, each with \texttt{$R(\texttt{VAR}) = \{1, 2, 3, 4\}$} as the range of allowed values. The test cases were executed on a machine with an Intel Core i7-6700 CPU with 3.40 GHz and 16 GiB RAM. The execution time of the file preparation, that is copying the files and converting all \texttt{\#if}-conditions, was measured.

\begin{figure}[ht]
	\centering
	\includegraphics[width=\columnwidth]{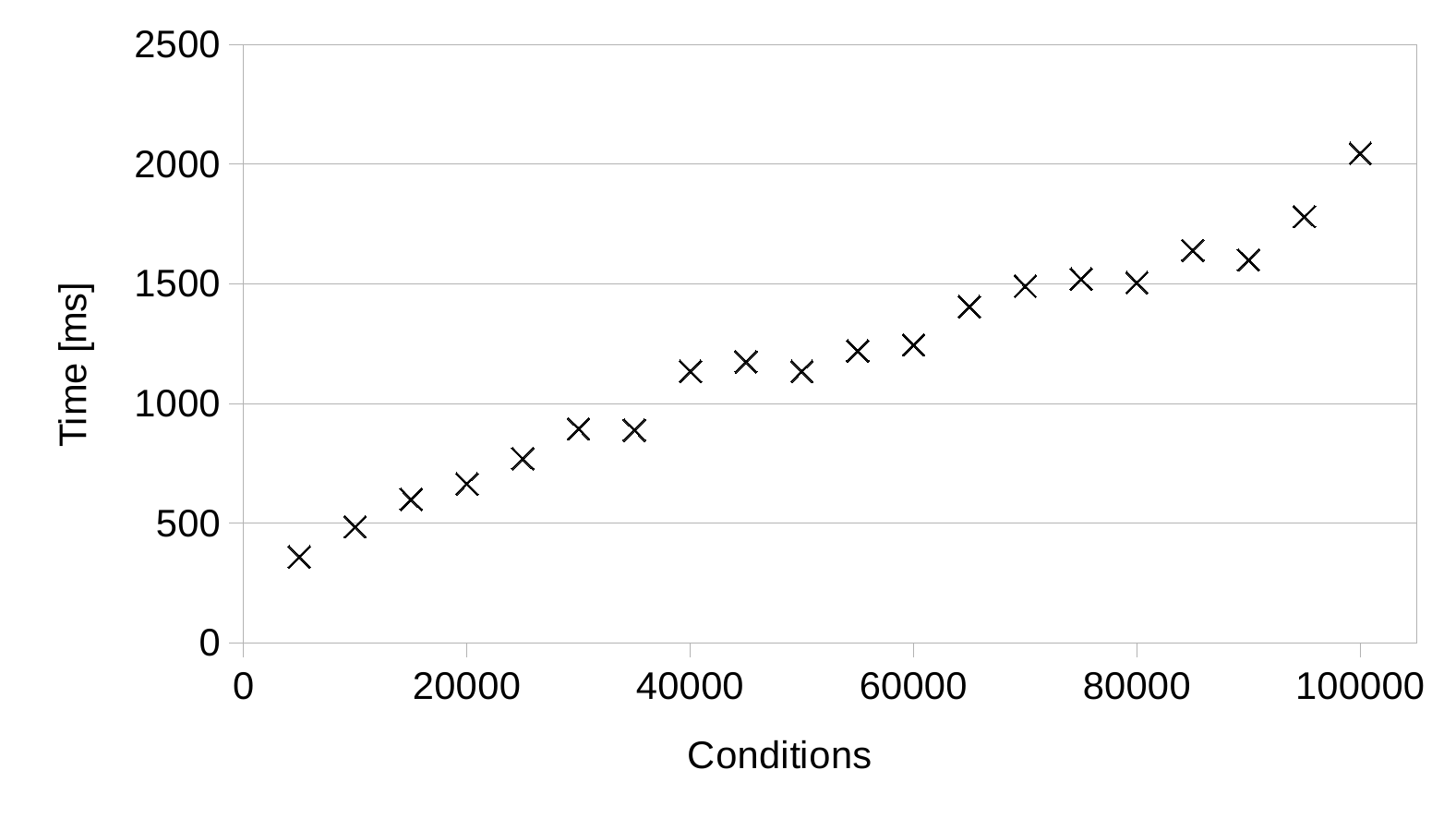}
	\caption{Runtime with varying number of conditions}
	\label{fig:time_per_conditions}
\end{figure}

Figure~\ref{fig:time_per_conditions} shows how the total number of \texttt{\#if}-conditions in the C source files relates to the execution time of the tool. For this test series, the number of \texttt{\#if}-conditions per C source file is increased in steps of 50, starting from 50 and up to 1000 conditions per file. Figure~\ref{fig:time_per_conditions} shows that the execution time grows linearly with the number of conditions to process. This means, that when analyzing a whole product line, the overhead of the preparation will also grow linearly with the size of the product line. 

\begin{figure}[ht]
	\centering
	\includegraphics[width=\columnwidth]{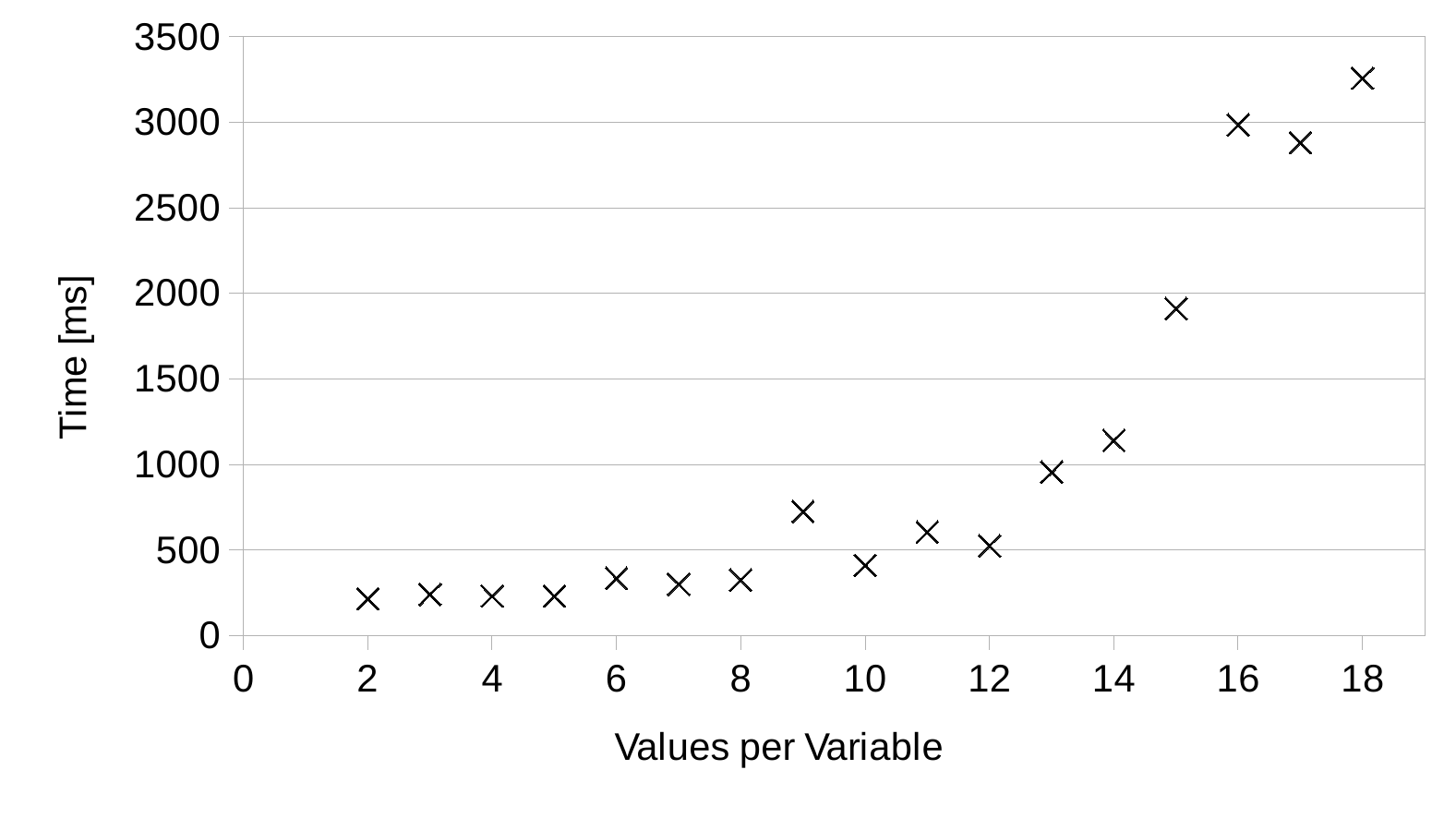}
	\caption{Runtime with varying ranges per variable}
	\label{fig:time_per_values}
\end{figure}

Figure \ref{fig:time_per_values} shows how the number of allowed values per variable impacts the execution time. For this test series, the size of the range of allowed values per variable was gradually increased, starting from 2 and up to 18. Additionally, the upper limit for the number of value combinations to consider (see Section~\ref{sec:limitations}) was removed. When an integer operator has variables on both sides, our approach considers all combinations of the allowed values for both variables. This leads to a quadratic growth of combinations, that is visible in Figure~\ref{fig:time_per_values}. However, when the upper limit for combinations to consider is not removed, the execution time stays below 500~ms for all test cases shown in this Figure. Limiting the number of combinations to consider produces less accurate results (see Section~\ref{sec:limitations}), but it mitigates the (potential) performance problem visible in Figure~\ref{fig:time_per_values}.

The implementation of the approach presented in this paper has also been used in practice in the analysis of the Bosch PS-EC product line \cite{BoschPaper}. Our tool converted the integer-based C-preprocessor conditions to propositional logic. This allowed existing SAT-based analyses to be used on the product line, without any modification to the existing analysis tools. The execution time of our preparation tool was not significant, compared to the execution time required for the following analysis steps. The resulting replacement conditions created by our approach allowed the following analysis steps to create meaningful results.

\section{Summary \secSize{0.75p}}
\label{sec:summary}

In this paper, we developed an approach to convert integer-based variability conditions to propositional logic. The original conditions are defined using the C-preprocessor in source code files. The variability variables used in the conditions hold integer values and are restricted to a (usually small) range of allowed values. Our approach converts all the conditions found in the source files, and replaces them with the created propositional formulas.

The goal of our approach is to easily use SAT-solvers even on integer-based product lines. This enables the usage of a number of existing, SAT-based tools and approaches without any modification to them. Thus, our approach ensures that the propositional formulas, that replace the integer-based one, are mostly equal to the original conditions with respect to satisfiability.

Section \ref{sec:concept} described the general concept of our approach: Boolean variables are introduced for each allowed value of the integer variables. This is viable, because the range of allowed values of a variable is usually small. Our approach then converts the integer-based conditions using these Boolean variables. It calculates for each integer (sub-)expression, which combination of allowed values fulfills it. A propositional formula, which reflects this combination of values that fulfill the expression, is used as the replacement.

Section \ref{sec:realization} explained how this is implemented. A set of rules is applied on the abstract syntax tree, to evaluate the integer-parts bottom-up. This evaluation keeps track of all allowed values of the integer variables at once. When a comparison operator is reached, a propositional formula can be constructured that specifies which combinations fulfill this comparison.

The approach also has a few limitations, as described in Section~\ref{sec:limitations}. Most importantly, the handling of unrestricted integer variables is not exact. Unrestricted integer variables are variables that have no restrictions on the allowed values, or have a very large range of allowed values. For these variables, our approach does not evaluate each possible value individually. Instead, only a single Boolean variable is used, that specifies whether the variable is set to \textit{any} value, or whether it is left undefined. This retains some useful variability information, although it is not as exact as the per-value analysis. In practice, this inexact approach still leads to reasonable results in the further analyses.

Our approach is implemented as an open-source plugin for the KernelHaven analysis framework \cite{KernelHavenPaper, nonbooleanutils_tool}. This implementation has been used in practice to analyze the Bosch PS-EC product line \cite{BoschPaper}.

Future work on integer-based product lines can utilize the approach presented in this paper to efficiently re-use existing, SAT-based analyses. Additionally, the approach presented here can be refined in the future. We already experiment with a heuristic to find intervals of unrestricted integer variables, that can be treated as a single value. These equivalence classes would allow a per-value analysis of unrestricted variables in the approach presented here, and would thus improve the quality of the results.

\vspace*{1em}
\begin{acks}
This work is partially supported by the ITEA3 project $\text{REVaMP}^2$, funded by the \grantsponsor{01IS16042H}{BMBF (German Ministry of Research and Education)}{https://www.bmbf.de/} under grant \grantnum{01IS16042H}{01IS16042H}. Any opinions expressed herein are solely by the authors and not by the BMBF.
\end{acks}
\vspace*{10em}

\flushcolsend


\bibliographystyle{ACM-Reference-Format}
\bibliography{literature}

\end{document}